\documentstyle[epsf,multicol_v1.1j,aps,prd]{revtex}
\begin{document}
\thispagestyle{empty}
{\baselineskip0pt
\leftline{\large\baselineskip16pt\sl\vbox to0pt{\hbox{Department of Physics} 
               \hbox{Kyoto University}\vss}}
\rightline{\large\baselineskip16pt\rm\vbox to20pt{\hbox{KUNS-1323}
           \hbox{OU-TAP-15} 
           \hbox{YITP/K-1100} 
           \hbox{astro-ph/???????}
           \hbox{February 1995}
\vss}}%
}
\vskip3cm
\begin{center}{\large The Hubble Parameter in Void Universe}
\end{center}
\begin{center}{\em -Effect of the Peculiar Velocity-}
\end{center}
\begin{center}
 {\large $^{1}$Ken-ichi Nakao, $^{2}$Naoteru Gouda, $^{1}$Takeshi Chiba, 
$^{2}$Satoru Ikeuchi}\\
{\large $^{3}$Takashi Nakamura and $^{2}$Masaru Shibata} \\
{\em $^{1}$Department of Physics, Kyoto University} \\
{\em Sakyo-ku, Kyoto 606-01, Japan } \\
{\em $^{2}$Department of Earth and Space Science, Faculty of} \\
{\em Science, Osaka University, Toyonaka, Osaka 560, Japan} \\
{\em $^{3}$Yukawa Institute for Theoretical Physics} \\
{\em Kyoto University, Sakyo-ku} \\
{\em Kyoto 606-01,~Japan}
\end{center}
\begin{abstract}
We investigate the distance-redshift relation in the simple 
void model. As discussed by Moffat and Tatarski, if the observer stays 
at the center of the void, the observed Hubble parameter is not so 
different from the background Hubble parameter.  
However, if the position of observer is off center of the void, 
we must consider the peculiar velocity correction which is measured by
the observed dipole anisotropy of cosmic microwave background. 
This peculiar velocity correction 
for the redshift is crucial to determine the Hubble parameter and 
we shall discuss this effect. Further the results of 
Turner et al by the N-body simulation will be also considered.
\pacs{PACS numbers: ????????}
\end{abstract}

\multicols{2}
Recent observation suggests that Hubble parameter is large one, that is,
$80 \pm 17$km/sec/Mpc (Freedman et al. 1994).
The low Hubble universe, however, is favored 
since the small value of Hubble parameter is consistent 
with almost all observations except for that of 
the Hubble parameter itself (Bartlett et al. 1994). 
One of the theoretical bases for the possibility of smaller Hubble 
parameter than that determined by local observation is given by 
Turner, Cen and Ostriker (Turner et al. 1992). 
They performed very large scale N-body simulations 
and constructed the ensemble of universe filled with the galaxies 
which, roughly speaking, are defined by density peaks of 
collisionless 
particles. 
Then, one of those galaxies is identified as ``our galaxy" and 
they investigate the relation between the distance of the other galaxies 
from our galaxy and the relative velocity with the correction about the 
peculiar velocity only of our galaxy. 
Their result suggests that the Hubble parameter determined 
by such observations has the scale dependent variance. In order to obtain 
the correct Hubble parameter, we need the observation of 
galaxies over the very wide region. 

On the other hand, Moffat and Tatarski considered the void universe 
in which the observer is assumed to be at the center of void and 
investigated the effect of the void on the Hubble parameter determined 
through 
the redshift and distance relation(Moffat $\&$ Tatarski 1994). 
Their result reveals that 
when the observer is at the center of void, the Hubble parameter 
is not so different from the true value as long as 
the observed region is smaller than the curvature radius within 
the void. This seems to  
contradict with the results of Turner et al. 

In this paper, we investigate the void universe, but shall not 
restrict the position of the observer to be the center of the void. 
Our void model is more simplified one than that of Moffat and Tatarski, 
but will clarify the effect of the inhomogeneities 
on the observation of the Hubble parameter. 
We assume that the inside of the void is approximated by the 
Friedmann-Robertson-Walker (FRW) 
universe with the present density parameter $\Omega_{0}<1$ while 
the outside of the void is also the FRW universe but with $\Omega_{0}=1$. 
The boundary of the void can be ignored as long as we exist within the 
void and observe only inside of that. Here we will assume such a 
situation. 
Further we assume that the age of both inside and outside  of 
the void is the same and hence the time coordinate is common cosmic time $t$ to 
both the inside and outside of the void. 
This assumption corresponds to the fact that the void structure comes from 
purely growing mode of the initial density perturbation 
since the density contrast between the inside and 
outside of the void vanishes as $t \rightarrow 0$, i.e., at the 
initial singularity. 

The metric within the void is written as 
\begin{equation}
ds^{2}=-dt^{2}+{a_{v}^{2}(t) \over 1+(R_{v}/R_{c})^{2}}dR_{v}^{2}
+a_{v}^{2}(t)R_{v}^{2}dS^{2}, 
\end{equation}
where $R_{c}$ is the comoving curvature radius and $dS^{2}=d\theta^{2}
+\sin^{2}\theta d\varphi^{2}$ is the line element on the unit sphere. 
We should note that the center of the void agrees with the origin $R_{v}=0$ 
and hence, as for the time coordinate $t$, 
$dS^{2}$ is common to the inside and outside of the void. 
As is well known, the scale factor $a_{v}$ is given as the parametric form 
by the conformal time $\eta$, 
\begin{eqnarray}
{a_{v}\over a_{v0}}&=&{\Omega_{v0} \over 2(1-\Omega_{v0})}(\cosh\eta-1), \\
H_{v0}t&=&{\Omega_{v0} \over 2(1-\Omega_{v0})^{3/2}}(\sinh\eta-\eta),
\end{eqnarray}
where $H_{v0}$,  $a_{v0}$  and $\Omega_{v0}$ are, respectively, the 
present Hubble parameter, the present scale factor and 
the present value of the density parameter, within the void. 

On the other hand, we assume that the outside of the void is 
the flat FRW universe and hence its metric outside the void is 
given by
\begin{equation}
ds^{2}=-dt^{2}+a_{b}^{2}(t)(dR_{b}^{2}+R_{b}^{2}dS^{2}), 
\end{equation}
and the scale factor $a_{b}$ is written as 
\begin{equation}
{a_{b} \over a_{b0}}=\Bigl({9 \over 4}H_{b0}^{2}t^{2}\Bigr)^{1/3}, 
\end{equation}
where $a_{b0}$ and $H_{b0}$ are, respectively, the present scale factor 
and the present Hubble parameter, outside the void. 

As discussed by Bartlett et al. (1994), 
the ratio, $H_{v0}/H_{b0}$, varies over the range $3/2$ to $1$ as 
$\Omega_{v0}$ varies form $0$ to $1$. Hence the maximum Hubble parameter 
within the void is at most $3/2$ times the background Hubble parameter 
$H_{b0}$. However, it should be noted that $H_{v0}$ is not observed 
directly. The observed Hubble parameter is determined through the 
relation between the distance and redshift with the correction 
about the peculiar velocity of both the observer and the observed source. 

Here we define the peculiar velocity which is crucial to 
estimate the true redshift. 
Assuming that the cosmic microwave background (CMB) radiation 
is homogeneous and isotropic, the peculiar velocity is defined 
as that against the frame in which the CMB is observed 
to be isotropic. 
Since the comoving observer outside the void
just observes the isotropic CMB, 
we first define the new radial coordinate ${\tilde R}$ for 
inside of the void as $a_{v}(t)R_{v}=a_{b}(t){\tilde R}$. 
It should be noted that the observer along ${\tilde R}$=constant curve 
looks just isotropic CMB. 
The transformation matrix is given by
\begin{eqnarray}
d{\tilde t}&=&dt,\\
d{\tilde R}&=&{a_{v} \over a_{b}}(H_{v}-H_{b})R_{v}dt+{a_{v} 
\over a_{b}}dR_{v}, \\
d{\tilde S}^{2}&=&dS^{2}.
\end{eqnarray}
In the original coordinate (1), the comoving observer and comoving 
observed source move along $R_{v}=$constant lines 
and hence the components of those 
4-velocities are given by the common $u^{\mu}=(1,0,0,0)$. 
On the other hand, in the above tilde coordinate 
system, the components are given by
\begin{eqnarray}
u^{\tilde t}&=&{\partial {\tilde t} \over \partial t} u^{t}=1, \\
u^{\tilde R}&=&{\partial {\tilde R} \over \partial t} u^{t}
={a_{v} \over a_{b}}(H_{v}-H_{b})R_{v}, \\
u^{\tilde \theta}&=&0=u^{\tilde \varphi}, 
\end{eqnarray}
and the radial component $u^{\tilde R}$ corresponds to the peculiar 
velocity of the comoving observer in the void. 

In order to obtain the relation between the redshift and the 
distance, it is sufficient to approximate  the 
light ray by a null geodesic, i.e., to treat the propagation of 
light ray by the geometric optics (Misner, Thorn $\&$ Wheeler 1973). 
By virtue of the spherical symmetry of this system, without loss of 
generality, we focus only on the null geodesic within the equatorial 
plane $\theta=\pi/2$. The solution for the null geodesic tangent $k^{\mu}$ 
is then given by 
\begin{eqnarray}
k^{t}&=&{a_{v0} \over a_{v}(t)}\omega_{v0},\\
k^{R_{v}}&=&\pm{a_{v0} \over a_{v}^{2}(t)}
      \sqrt{[1+(R_{v}/R_{c})^{2}][\omega_{v0}^{2}
      -(L_{0}/R_{v})^{2}]}, \\
k^{\varphi}&=&{a_{v0}L_{0} \over a_{v}^{2}(t)R_{v}^{2}}, 
\end{eqnarray}
and $k^{\theta}=0$. The radial trajectory of the null geodesic 
is obtained as  
\begin{equation}
R_{v}=R_{k}(\eta) \equiv R_{c}\sqrt{F^{2}(\eta)-1} ,  
\end{equation}
with 
\begin{eqnarray}
F(\eta)&=&\sqrt{ 1+\Bigl({L_{v0} \over \omega_{v0}R_{c}}\Bigr)^{2}}
\cosh\Bigl[\cosh^{-1}
\Bigl\{\sqrt{1+\Bigl({R_{v0} \over R_{c}}\Bigr)^{2}} \nonumber \\
&/&\sqrt{1+\Bigl({L_{v0} \over \omega_{v0}R_{c}}\Bigr)^{2}}\Bigr\}
\pm(\eta-\eta_{0})\Bigr],
\end{eqnarray}
where $R_{v0}$, $L_{v0}$ and $\omega_{v0}$ are the integration constants and 
$\eta_{0}$ is the present conformal time. 
It should be noted that, at $\eta=\eta_{0}$, $(R_{v}, \varphi)=(R_{v0},0)$ 
and this corresponds to the position of the 
comoving observer at the moment of observation. 
$L_{v0}$ is the conserved 
angular momentum of the light ray while, $\omega_{v0}$ is the angular 
frequency of that for the comoving observer. 
Together with $\omega_{v0}$, $L_{v0}$ determines 
the angle $\theta_{k}$ between the radial direction and the propagation  
direction of the light ray as, (see Fig.1), 
\begin{equation}
 \cos\theta_{k}=\mp
\sqrt{1-\Bigl({L_{v0} \over \omega_{v0}R_{v0}}\Bigr)^{2}}. 
\end{equation}

Next, we consider the effect of the peculiar velocity on the 
angular frequency of the light ray. 
The comoving observer (comoving observed source) 
detects (emits) the light ray $k^{\mu}$ with the angular 
frequency, $\omega_{v} \equiv -k_{\mu}u^{\mu}=-k_{t}$ 
On the other hand, the observer and observed source 
moving along ${\tilde R}=$constant curve, 
have 4-velocity $w^{\tilde \mu}=(1,0,0,0)$ in the tilde coordinate 
and hence the angular frequency for those is given by 
\begin{eqnarray}
   \omega_{c}&\equiv& -k_{\tilde \mu}w^{\tilde \mu}=-k_{\tilde t}
   =\omega_{v}+k_{\tilde R}u^{\tilde R} \nonumber \\
   &=&\omega_{v}+(H_{v}-H_{b})R_{v}k_{R_{v}}.
\end{eqnarray}
It should be noted that $\omega_{c}$ corresponds to the angular frequency 
with the correction for the peculiar velocity. 
Observationally, we can consider the effect only of our own 
peculiar velocity and hence hereafter we focus on the 
quantities with the correction about the peculiar velocity only of the 
observer and those without any corrections for the peculiar velocity. 
Then we define the following two kinds of the redshift as
\begin{equation}
  z={\omega_{v} \over \omega_{v0}}-1,~~~~ {\rm and}~~~~
  z_{co}={\omega_{v} \over \omega_{co}}-1,
\end{equation}
where $\omega_{v}$ is the angular frequency of the light ray at the 
observed source while $\omega_{co}$ is given by 
\begin{equation}
\omega_{co}=\omega_{v0}+(H_{v0}-H_{b0})R_{v0}k_{R_{v}}(\eta_{0}).
\end{equation}
Hence, $z$ is the bear observed redshift and $z_{co}$ is the redshift 
with the correction about the peculiar velocity only of the 
observer. 

We shall employ the luminosity distance $D_{L}$ as the 
distance measure between the observer and observed source. 
Here, the luminosity distance $D_{L}$ is given by 
well-known relation in the FRW universe with (1) as 
\begin{equation}
  D_{L}={1 \over H_{v0}q_{v0}^{2}}[zq_{v0}+(q_{v0}-1)
       (-1+\sqrt{2q_{v0}z+1})], 
\end{equation}
where $q_{v0}=\Omega_{v0}/2$. 
It should be noted that the luminosity distance $D_{L}$ is 
just the observed quantity which is determined by, for example, 
Tully-Fisher relation. 
Then, using $D_{L}$, we define the observed Hubble parameter $H_{co}$ 
with the correction for the peculiar velocity only of the observer  
with the assumption that observers regard their own universe 
as the flat FRW space-time,  
\begin{equation}
 H_{co}={2 \over D_{L}}[z_{co}+1-\sqrt{z_{co}+1}].
\end{equation}
In fact, we can measure $H_{co}$ instead of $H_{b0}$ in the real observations.
In Fig.2, $H_{co}$ is depicted for $\theta_{k}=0$, $\pi/2$ and $\pi$. 
In this figure, the density parameter inside the void, $\Omega_{v0}$ is 
equal to $0.1$ and the radial position of the observer 
is fixed as $a_{v0}R_{v0}=1 \times 10^{-2}H_{b0}^{-1} \sim 30h^{-1}_{b}$Mpc. 

We find that, for $H_{v0}D_{L} \ll 1$, $H_{co}$  strongly depends on the 
observed direction along which the light ray propagates. 
This comes from the wrong peculiar velocity correction and 
it should be noted that the Hubble parameter defined 
by Turner et al. is a volume average of just $H_{co}$. 

To understand the direction dependence of $H_{co}$, we 
investigate that only for $H_{v0}D_{L}\ll 1$. 
In this case, $H_{co} \sim z_{co}/D_{L} 
\sim H_{v0}(z_{co}/z)$ and, assuming the case of 
$\Omega_{v0} = 0.1 $, we obtain $H_{v0}/H_{b0}-1 \sim 0.35$. 
Further, the distance, $a_{v0}R_{v0}$, of the observer from the center 
of the void is assumed to be less than about $100h_{b}^{-1}$Mpc, i.e.,  
$a_{v0}H_{b0}R_{v0} < 3 \times 10^{-2} \ll 1$. 
Hence, we obtain 
\begin{eqnarray}
z_{co} & \sim & H_{v0}D_{L} - 10^{-2} 
              \times{1 \over \omega_{v0}} \nonumber \\
       & \times & (H_{v0}D_{L}+1) k_{R_{v}}(\eta_{0})
              \Bigr({R_{v0} \over 100{\rm Mpc}}\Bigr).
\end{eqnarray}
Since $a_{v0}R_{c}=H_{v0}^{-1}(1-\Omega_{v0})^{-1/2} \sim H_{v0}^{-1}$, 
$R_{v0}/R_{c}$ is much less than unity and hence we can see that 
$k_{R_{v}}(\eta_{0}) \sim -\omega_{v0}\cos \theta_{k}$. 
Then we get 
\begin{equation}
  {H_{co} \over H_{b0}} \sim {H_{vo} \over H_{b0}}
   + 10^{-2}  {1\over H_{v0}D_{L}}
   \cos\theta_{k} \Bigr({R_{v0} \over 100{\rm Mpc}}\Bigr).
\end{equation}
From the above equation, when the distance of the 
observer from the center of void is $30h^{-1}_{b}$Mpc 
and when such an observer 
looks to the direction $\theta_{k}=0$ and the observed distance 
is $ D_{L} = 3 \times 10^{-3} H_{v0}^{-1} \sim 7h_{b}^{-1}$Mpc, 
the observer may estimate 
$H_{co}$ to be factor two times larger than $H_{b0}$. 
On the other hand, if that observer looks to the opposite 
direction 
$\theta_{k}=\pi$, the observer may obtain almost vanishing $H_{co}$. 
This is just the dipole anisotropy due to the wrong correction 
for the peculiar velocity. 

Here we shall consider the relation between our simple void model and 
the results by Turner et al. 
In our case, the averaged $H_{co}$ agrees with $H_{v0}$ as 
\begin{equation}
   <H_{co}>={1 \over \pi}\int_{0}^{\pi} d\theta_{k}H_{co}=H_{v0}. 
\end{equation}
It should be noted that we assume the uniform distribution of observed 
source, i.e., galaxy when we perform the above averaging. 
However, in the N-body simulation, the ``galaxy" is not uniformly 
distributed in contrast with our model 
and the integral of the second term in R.H.S. of Eq.(24) may remain. 
Fig.1 shows an example in which the number of galaxies on the direction 
$\theta_{k}=0$ is larger than that on $\theta_{k}=\pi$ direction. 
In such a case, the averaged $H_{co}$ is greater than $H_{v0}$. 
Therefore it may be a reason 
why the variance of the Hubble parameter depends on the
scale of the observational regions and there appears the large 
variance of 
the Hubble parameter 
in the small scale observation in the results of Turner et al. 
Of course, in order to confirm this expectation, the detailed investigation 
by the N-body simulation is needed (Gouda et al. 1995). 

From the observational point of view, if the variance of Hubble parameter 
comes from the dipole anisotropy such as above, it is important 
to confirm the isotropy of Hubble parameter. Lauer and Postman 
reported the highly isotropic Hubble parameter by the 
rather large scale observation $z \leq 0.05$ (Lauer $\&$ Postman 1992). 
Hence, even if we stay in the void, we are near the center of that. 
In the case that we stay near the center of the void, 
the observed Hubble parameter is $H_{v0}$ and this varies 
over the range $H_{b0}$ to $1.5H_{b0}$. 
Since this is not so large variance, we can find almost the same 
Hubble parameter as the background one. 
Of course, our model is too simple and more complicated situations 
may be imagined, which makes us to fail the true Hubble parameter. 
Hence further theoretical investigation should be continued and 
deeper observation over whole direction in the sky is very important. 

Finally, we should comment on the effect of the void on the anisotropy 
of CMB. Here we shall assume that the CMB is completely isotropic at the 
last scattering surface and that the anisotropy is caused only by the 
effect of one void. The dipole anisotropy of CMB is about 
$v/c$ where $v$ is the peculiar velocity of the comoving 
observer in the void and it is given roughly as $(H_{v0}-H_{b0})a_{v0}
R_{v0}$ 
by Eq.(10). If the density parameter inside the void is nearly zero, 
we obtain $v \sim 1.5\times10^{3}(a_{v0}R_{v0}/100h^{-1}_{b}{\rm Mpc})$km/sec. 
Assuming that the observed dipole anisotropy comes form 
the peculiar velocity of our local group, 
that is estimated as about $600$km/sec (Smoot et al 1991). 
If we live in such a void, then 
our position is $10h^{-1}_{b}$Mpc apart from the center of the void. 
However our void considered here is nothing but a toy model and it 
should not be seriously considered. 

The rather serious subject is the quadrupole or higher multi-pole 
anisotropies which come from the gravitational redshift. 
We consider the situation that the size of the 
void is sufficiently smaller than the horizon scale $L$ of the background 
flat FRW universe and hence the Newtonian approximation is 
applicable. 
In this case, the metric is written as
\begin{equation}
 ds^{2}=-(1-2U)dt^{2}+a_{b}^{2}(t)(1+2U)(dR^{2}+R^{2}dS^{2}), 
\end{equation}
where $|U| \ll 1$.
Further we assume the step-function-like density configuration, 
\begin{equation}
  \rho=\left\{\begin{array}{ll}
              \rho_{v}(t) & \mbox{$R<R_{void}$} \\
              \rho_{b}(t) & \mbox{otherwise}
              \end{array}
       \right. 
\end{equation}
where $\rho_{b}$ corresponds to the critical density. 
Then the Newton potential $U$ inside the void $R<R_{void}$ is obtained as
\begin{equation}
  U=2\pi\delta\rho \ell^{2} -{2\pi \over 3}\delta\rho (a_{b}R)^{2}, 
\end{equation}
where $\ell \equiv a_{b}R_{void}$. Here, since we consider the case in which 
$\delta \rho \sim -\rho_{b} \sim -H_{b}^{2} = -L^{-2}$, we 
see that $\delta\rho \ell^{2} \sim \kappa^{2} \equiv (\ell/L)^{2} \ll 1$. 
Thus we can roughly estimate the Newtonian potential as 
$U \sim \kappa^{2} - \kappa^{2}(a_{b}R / \ell)^{2}$, 
$\partial_{t}U \sim H_{b}U$ and 
$\partial_{r}U \sim \kappa^{2}( a_{b}/\ell)^{2}R$. 
Here we shall estimate the Sachs-Wolfe effect on the CMB by 
the above Newtonian potential. 
The anisotropy of CMB is expressed by the integrated brightness 
temperature perturbation $\Theta$ and the equation for $\Theta$ is 
written as 
\begin{equation}
  {d \over dt}(\Theta-U) \equiv 
  \Bigl(\partial_{t}+{\gamma^{i} \over a_{b}}\partial_{i}\Bigr)(\Theta-U)
  =-2\partial_{t}U, 
\end{equation}
where $\gamma^{i}$ is the direction cosine of the photon 
(Kodama $\&$ Sasaki 1986).
Then, the difference between the two opposite 
radial directions is roughly estimated as 
\begin{equation}
   {\Delta T \over T} 
=2\Bigl(\int dt \partial_{t}U|_{\theta_{k}=0}-
 \int dt \partial_{t}U|_{\theta_{k}=\pi}\Bigr) \sim 
 \kappa^{3}\Bigl({R_{o} \over \ell}\Bigr), 
\end{equation}
where $R_{o}$ denotes the radial position of the observer and the 
integration is performed along the path of the light ray. 
It should be noted that the above result is consistent with the 
analysis by Meszaros (1994) for the case that the position of 
the observer is outside of the void.  
Hence if we live in the 
$100h^{-1}_{b}$Mpc scale void, since $\kappa^{3} \sim 4 \times 10^{-5}$, 
the higher multi-pole anisotropy of CMB by the such a void does 
not conflict with $COBE$ results (Smoot et al. 1992). 
However, this estimate is so rough that we need more detailed 
investigation and this is in progress. 


\vskip 0.3in
\centerline{ACKNOWLEDGMENTS}
\vskip 0.05in
We would thank to H. Sato and M. Sasaki for their useful
discussion. KN would like to thank T. Tanaka for his crucial 
suggestion on the peculiar velocity. 

\endmulticols
\vskip1cm
\begin{center}
{\large REFERENCES}
\end{center}
\multicols{2}
\noindent
Freedman W.L. et al., 1994, Nature, {\bf 371}, 757 

\noindent
Bartlett J.G., Blanchard A., Silk J., Turner M., 1994, 
FERMILAB-Pub-94/173-A (1994)

\noindent
Turner E.L., Cen R., Ostriker J.P., 1992, Astron. J., {\bf 103}, 1427

\noindent
Moffat J.W., Tatarski D.C., 1994, preprint, UTPT-94-19

\noindent
Misner C.W., Thorne K.S., Wheeler J.A., 1973, {\it Gravitation}(Freeman)

\noindent
Lauer T.R., Postman M., 1992, ApJ, {\bf 400}, L47

\noindent
Smoot G. F. et al, 1991, ApJ, {\bf 371}, L1

\noindent
Kodama H., Sasaki M., 1986, Intern. J. Mod. Phys., {\bf A1}, 256

\noindent
Meszaros A., 1994, ApJ, {\bf 423}, 19

\noindent
Gouda N. et al., 1995, in preparation

\noindent
Smoot G.F et al, 1992, ApJ, {\bf 396}, L1

\endmulticols

\begin{figure}
 \centerline{\hspace{-4cm}\epsfxsize13cm \epsfbox{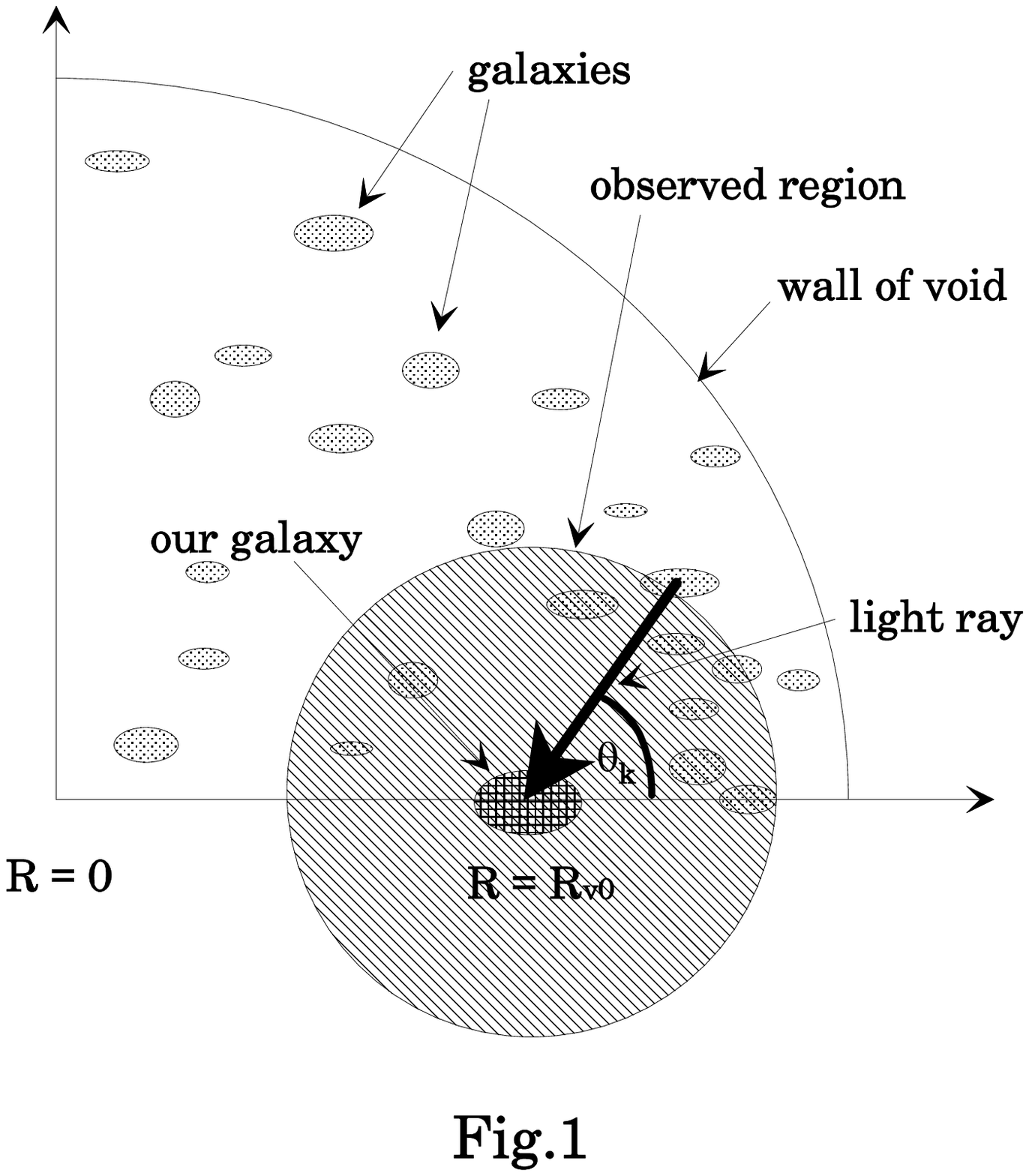}}
 \vspace*{-1.5cm}
 \centerline{\footnotesize 
The schematic diagram of the position of observer and 
the observed direction. The angle $\theta_{k}$ is defined in Eq.(17).
}
\end{figure}
\begin{figure}
 \centerline{\epsfxsize8cm \epsfbox{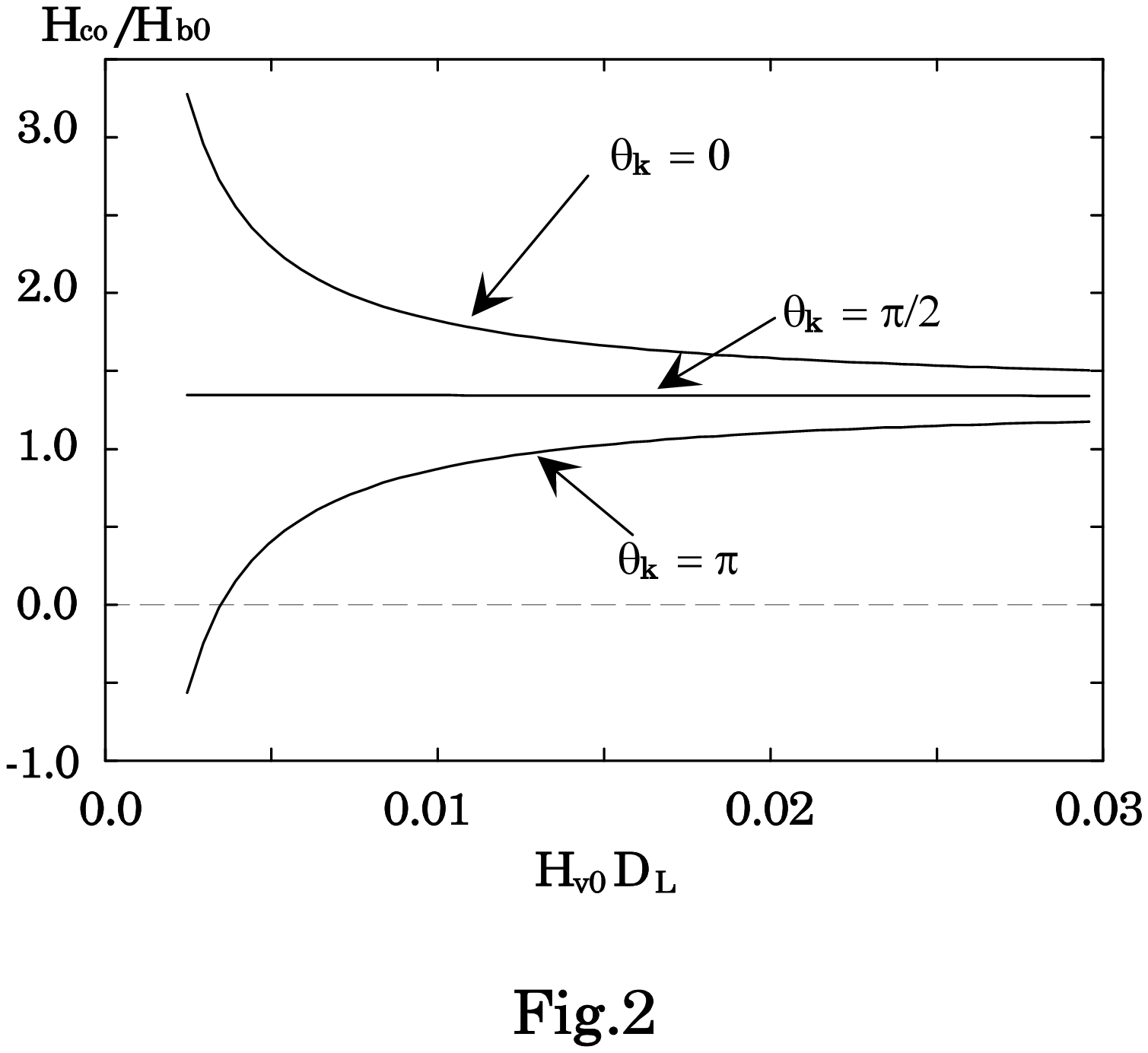}}
 \vspace*{5mm}
 \centerline{\begin{minipage}[t]{15cm}{\footnotesize
 The Hubble parameter $H_{co}$ with the correction for 
the peculiar velocity only of the observer is plotted against the 
luminosity distance $D_{L}$ for various direction. 
The density parameter $\Omega_{v0}$ within the void is $0.1$.
}\end{minipage}
}
\end{figure}

\end{document}